\documentclass[pre,twocolumn,twoside,aps,floatfix,longbibliography,natbib,nofootinbib,superscriptaddress]{revtex4-1}

\usepackage{pslatex,graphicx,dcolumn,bm,natbib,amssymb,amsmath,color,mathtools,mathrsfs,eufrak}

\usepackage[utf8]{inputenc}
\usepackage[T1]{fontenc}
\usepackage{hyperref}
\usepackage[english]{babel}
\usepackage{blindtext}
\usepackage{booktabs}

\hypersetup{
	colorlinks=true,
	linkcolor=blue,
	citecolor=green,
	filecolor=blue,
	urlcolor=blue
}

\usepackage{epstopdf}

\definecolor{cream}{RGB}{222,217,201}

\begin{document}

\title{Bridging the Gap Between Collective Motility and Epithelial-Mesenchymal Transitions through the Active Finite Voronoi Model}

\author{Junxiang Huang}
\affiliation{Department of Physics, Northeastern University, Boston, MA 02115, USA}
\author{Herbert Levine}
\affiliation{Department of Physics, Northeastern University, Boston, MA 02115, USA}
\affiliation{Center for Theoretical Biological Physics and Departments of Bioengineering, \\Northeastern University, Boston, MA 02115, USA}
\author{Dapeng Bi}
\affiliation{Department of Physics, Northeastern University, Boston, MA 02115, USA}

\begin{abstract}
We introduce an active version of the recently proposed finite Voronoi model of epithelial tissue. The resultant Active Finite Voronoi (AFV) model enables the study of both confluent and non-confluent geometries and transitions between them, in the presence of active cells. Our study identifies six distinct phases, characterized by aggregation-segregation, dynamical jamming-unjamming, and epithelial-mesenchymal transitions (EMT), thereby extending the behavior beyond that observed in previously studied vertex-based models. The AFV model with rich phase diagram provides a cohesive framework that unifies the well-observed progression to collective motility via unjamming with the intricate dynamics enabled by EMT. This approach should prove useful for challenges in developmental biology systems as well as the complex context of cancer metastasis. The simulation code is also provided at \hyperlink{https://github.com/jxhuangphys/Active_Finite_Voronoi_simulation}{\texttt{github.com/jxhuangphys/Active\_Finite\_Voronoi\_simulation}}.
\end{abstract}
\maketitle 


During development and remodeling, as well as during wound healing and invasion, the ability of cells to migrate collectively \cite{friedl2009collective,friedl2012classifying} is crucial for biological functions. 


Mesenchymal phenotypes~\cite{Kalluri_Weinberg_review_2009} exhibit distinct cellular behaviors during movement, characterized by features like polarized morphology and dynamic cytoskeletal changes. One notable trait is their tendency to exhibit lower confluency, resulting in fewer triple junctions. This reduced cell-cell contact allows mesenchymal cells to adapt to dynamic microenvironments more readily, facilitating their capacity for individual or collective migration, invasive behavior, and responsiveness to chemical and mechanical cues. In contrast, epithelial tissues maintain a tightly packed and organized structure with extensive intercellular contacts such as adherens junctions, tight junctions, and desmosomes~\cite{GIEPMANS2009820}. These junctions create a cohesive barrier, essential for functions like absorption, secretion, and providing a protective layer. Unlike mesenchymal cells, epithelial cells typically find themselves ``caged'' by their neighbors, an arrangement that shares a striking resemblance with glassy materials~\cite{berthier_biroli_review}. Despite these constraints, cells within an epithelium can still move collectively as sheets rather than individually, relying on coordinated cell-cell interactions to maintain tissue integrity and function~\cite{friedl2009collective}. This fundamental distinction in cellular behavior underscores the diverse roles these tissue types play in physiological processes and brings forth numerous crucial challenges in the fields of active matter physics and non-equilibrium statistical mechanics~\cite{marchetti2013hydrodynamics,Hakim_2017,alert_trepat_reivew}.


Research over the past several years \cite{
Park_NMAT_2015,
Garcia2015,
delarue2016self,
Oswald2017,
malinverno2017endocytic,
atia2018geometric,
mongera2018fluid,
Ilina2020,
mitchel_ncomm_2020,
petridou2021rigidity,o2020irradiation} has demonstrated the importance of understanding epithelial motility through the paradigm of the jamming and unjamming transition. During the unjamming transition (UJT), an epithelial collective transforms from a jammed phase where cells behave solid-like, toward an unjammed phase where cells flow in a fluid-like manner. In both the jammed and unjammed phases, the cellular collective retains intact epithelial junctions and remain in a confluent state where there are no gaps between cells.

The UJT paradigm complements the existing mechanism of the epithelial-to-mesenchymal transition (EMT)~\cite{hay_ETM_1995,Kalluri_Weinberg_review_2009,lamouille2014molecular,campbell_casanova_EMT_review_2016,nieto2016emt,Francou_Anderson_review_2020}. During EMT, an epithelial cell progressively acquires mesenchymal characteristics that, in the limit of full EMT, lead to bulk dissociation and single-cell, dispersed, mesenchymal migration. This transition to a migratory state is defined by disruption of apico-basal polarity and cell-cell junctions. Graded changes along this axis define epithelial plasticity often described in terms of ‘partial EMT’ or ‘hybrid’ E/M states~\cite{jolly_hypbrid_EMT}. Therefore, understanding the interplay of UJT and EMT in the mechanics and organization of a multicellular collective is crucial.

In the past two decades, a number of computational approaches have been proposed to understand the emergence of collective behaviors in multicellular systems. In particular, a class of cell-based models known as vertex models has been proven effective for capturing epithelial mechanics~\cite{Nagai_PMB_2001,Farhadifar_CB_2007,Fletcher2014,Alt_VM_review_2017}. In the vertex model, each cell is represented as a deformable polygonal inclusion, with edges and vertices shared by neighboring cells. 
This class of models necessarily assumes that exactly three cells meet at any vertex in an epithelial tissue and no gaps exist between cells, i.e. the tissue exists at the confluent limit. In the most common version, the dynamical degrees of freedom are the vertex locations. On the other hand, in a confluent Voronoi model, a cell $i$ is parameterized by the corresponding cell center $\mathbf{r_i}$, and cell shapes are determined by Voronoi tessellation based on the cell centers $\{\mathbf{r_i}\}$.

Several previous studies have explored the connections between the aforementioned EMT and tissue-level unjamming~\cite{haeger2014cell, sadati2013collective, mitchel_ncomm_2020}. This is particularly important in the context of partial EMT, which has also been shown to give rise to collective (as opposed to individual) cell motility~\cite{oncotarget}. In general, it appears that these processes represent distinct modes of ``liquefaction'', but the precise relationship between them has not yet been fully investigated. Previous efforts based on confluent Vertex or Voronoi-based models have had difficulties dealing with this question, due to the fact that these models specifically aim at simulating epithelial cells. An ideal candidate model to address this open question would have the ability to have both an unjamming transition and EMT (or a combination of both) under suitable parameters.

In general, then, although confluent vertex models and Voronoi-based approaches have successfully revealed the density-independent rigidity transition in dense epithelial systems, their ability to accurately describe cell movements and tissue properties in a low cell density is limited. Inspired by this limitation, various efforts have proceeded towards devising different models to describe non-confluent biological tissues. One example is provided by the cellular Potts model \cite{cpm}, which has no difficulty in generating nonconfluent phases if the cell-medium energy is lower than cell-cell one. Next, a generalized version of the vertex model allows for a deformable free-shape connected boundary but cannot accommodate topology changes~\cite{barton2017active}. In the Subcellular Element model, each individual cell is composed of numbers of ``elements'' which have short-range viscoelastic interactions, resulting in adaptive cell shapes and intercellular spaces~\cite{sandersius2008modeling, basan}. Kim {\it et al} add intermediate vertices into vertex model to allow for more complex cell shapes, and further introduce extracellular spaces by simulating them as ``virtual cells''~\cite{kim2021embryonic}. Finally, Loewe \textit{et al.} used a multi-phase field model, in which cells are treated as deformable and overlapping active particles, to allow for the emergence of inter-cellular gaps~\cite{loewe2020solid}. Many of these recent models require more degrees of freedom and  therefore are significantly more complex than vertex-models.

In one recent effort to bridge the gap between confluent and non-confluent tissue mechanics models while maintaining the simplicity and ease of use of earlier vertex models, Teomy \textit{et al} extended the confluent Voronoi approach to create a Finite Voronoi (FV) model~\cite{Teomy_Kessler_Levine_2018}. Here, a maximal size is assumed for the cells, which then guarantees that the resulting tissue becomes non-confluent at a sufficiently low density. While there has been a thorough study of the static morphological and thermal fluctuations of the FV model, the lack of inclusion to date of active forcing means that the dynamical organization of the multicellular structure and possible collective motility have remained unexplored. 

In this work, we construct an Active Finite Voronoi (AFV) model by incorporating into the FV model self-propelled active forces~\cite{Bi_PRX_2016,huang_shear_2022}. We systematically explore the interplay of activity and intercellular mechanical interactions and comprehensively map out the different emergent phases. In addition to recapitulating the previously observed confluent unjamming/jamming behaviors, we discover that activity and cell-cell interactions can also drive the tissue to undergo an epithelial-mesenchymal transition. Interestingly, the model exhibits a rich set of epithelial and mesenchymal morphological and dynamical phases. In fact, we reveal six different phases defined by an aggregation-segregation transition, a dynamical jamming-unjamming transition, and an epithelial-mesenchymal transition. The existence of these phases and the transitions between them could potentially provide novel insight into recently observed tissue behavior.

\section{Model}

\begin{figure}[h]
\centering
 \includegraphics[width=0.5\textwidth]{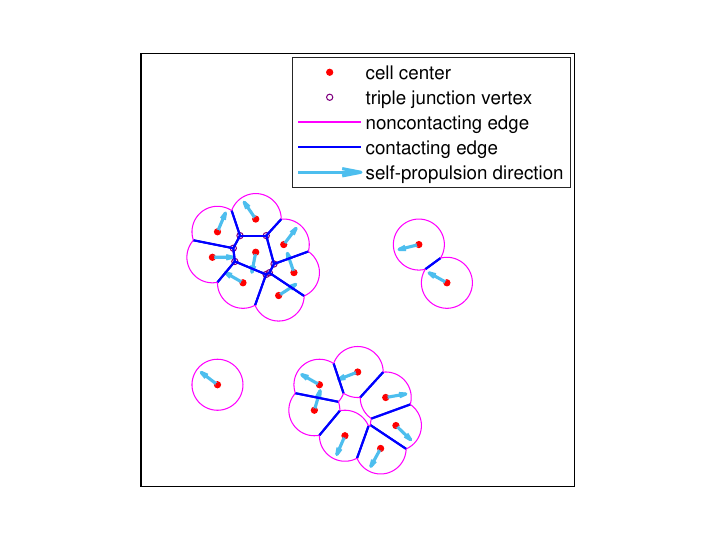}
	\caption{An example of a tissue snapshot from the AFV model illustrating its salient features. The red dots denote the cell centers. The blue lines represent contacting edges shared by two cells, and the pink curves represent non-contacting edges which are circular arcs with radius $l$. Triple junction vertices that connect three cells are indicated by hollow purple dots. The cyan arrows indicate the polarity directions of the self-propulsion force applied on the cell centers. Note that isolated cells are circular in shape.}
	\label{fig:sample}
\end{figure}

In a conventional Voronoi tessellation, space is partitioned based on the shortest distance between pairs of points. This allows Voronoi tessellations to tile all of space. At the edge of a cluster of points, the neighborhood of  points will extend to spatial infinity, giving rise to unbounded cell sizes. For this reason, conventional Voronoi-based models typically employ periodic boundary conditions. 
 In order to explore non-confluent regimes, following previous studies~\cite{Graner_Sawada_1993, Schaller_Meyer-Hermann_2005,Teomy_Kessler_Levine_2018}, the FV model augments a conventional Voronoi tessellation by introducing a length scale $l$, which sets the maximum neighborhood belonging to any point~\cite{Teomy_Kessler_Levine_2018}. In other words, every cell lies entirely within a circle of radius $l$ about the center. The resulting cell boundaries consist of both polygonal segments (contacting edges, shown as blue lines in Fig. \ref{fig:sample}) and circular arcs (non-contacting edges, shown as pink curves in Fig. \ref{fig:sample}): On one hand, cells separated by a distance less than $2l$ will still have contacting edges determined by the conventional Voronoi tessellation. On the other hand, for edges that are more than $l$ from cell centers, cell boundaries are replaced by circular arcs of radius $l$. This allows cell-unoccupied regions and intercellular gaps to arise naturally when two neighboring centers are more than $2l$ away. As in the standard Voronoi model, the cell center positions $\{\mathbf{r_i}\}$ are the dynamical degrees of freedom and the cellular structure are determined by the aforementioned combination of Voronoi tessellation and length scale $l$.

To incorporates the mechanics of the cell layer, we follow previous vertex-based model approaches~\cite{Farhadifar_CB_2007, Staple_PJE_2010} and write an energy function to, which captures cell-cell interactions and single cell-mechanics. Specifically,
%
\begin{align}
\tilde{E}
= \sum_{i=1}^N \big[ K_A (\tilde{A_i} - \tilde{A_0})^2 + K_P \tilde{P_i}^2 \big] +\lambda^{(c)}\sum 2l^{(c)} +\lambda^{(n)}\sum l^{(n)},
\label{energy_eq_first}
\end{align}
where $K_A$ and $K_P$ are the area and perimeter elastic moduli. $\{\tilde{A_i}\}$ and $\{\tilde{P_i}\}$ are cell areas and perimeters, and $A_0$ is the preferred area. The first term arises from resistance to cell volume change. The second term $K_P \tilde{P_i}^2$ results from energy cost of cortex deformation, due to the presence of proteins and other molecules such as actin and myosin that provide structural support to the cellular semi-permeable membrane~\cite{Farhadifar_CB_2007, Staple_PJE_2010}. $\lambda^{(c)}$ and $\lambda^{(n)}$ are correspondingly cortical tensions on contacting edges (cell-cell interfaces) and non-contacting edges (boundary edges). 

In epithelial cells, non-contacting edges tend to carry higher tension than contacting edges~\cite{Manning_Foty_Steinberg_Schoetz_2010}. This cortical tension difference origins from feedback between adhesion molecules and cytoskeletal dynamics. The normal projection of cortical tension on non-contacting edges (cell-unoccupied region interfaces) is balanced by cortical elasticity; the tension required for this force balance is lower for contacting edges (cell-cell interfaces) due to the contribution of adhesion from neighboring cells. The factor $2$ in the third term of Eq.~\ref{energy_eq_first} comes from the fact that each contacting edge is shared by two cells. Using the relationship $\sum_{i=1}^N \tilde{P_i}=\sum_{i=1}^N \tilde{L_i}^{(n)}+\sum 2l^{(c)}$, where $\tilde{L_i}^{(n)}$ is the total length of non-contacting edges in the $i$-th cell, the above equation can be simplified to
\begin{align}
\tilde{E}
&= \sum_{i=1}^N K_A (\tilde{A_i} - \tilde{A_0})^2 + K_P (\tilde{P_i} + \frac{\lambda^{(c)}}{2K_P})^2 +(\lambda^{(n)}-\lambda^{(c)}) \tilde{L_i}^{(n)} - (\frac{\lambda^{(c)}}{2K_P})^2,
\end{align}
where the last term is a constant and can be dropped. This equation can be further simplified in two ways: On one hand, $l$ can be used as the length unit in the system in order to non-dimensionalize the perimeter and area quantities, i.e. let $P_i=\tilde{P_i}/l$,
$L_i^{(n)}=\tilde{L_i}^{(n)}/l$, $A_i=\tilde{A_i}/l^2$, and $A_0=\tilde{A_0}/l^2$ be all dimensionless. We also introduce $P_0=-\frac{\lambda^{(c)}}{2K_Pl}$ as the dimensionless preferred cell perimeter, which indicates the relative strength of extensile tension on contacting edges versus elastic contractile tension. A high $P_0$ value may result from a high cortical tension on the edge, i.e. high $|\lambda^{(c)}|$, which is typical in fluid-like epithelial states; or it may reflect a weak membrane elasticity, i.e. small $K_P$, which is common in mesenchymal states. On the other hand, $K_A$ is extracted from the tensions and elastic coefficients, which gives $k_P=K_P/K_A$, along with $\Lambda=(\lambda^{(n)}-\lambda^{(c)})/K_A$, the normalized tension difference coefficient between contacting edges and non-contacting edges. A large $\Lambda$ value will encourage cells to form cell-cell interfaces with neighbors and eliminate intercellular gaps in order to reduce the total length of non-contacting edges. The inclusion of $\Lambda$ is similar to the inclusions of the cell-medium interaction in the cellular Potts models~\cite{mukherjee2021cluster}. With these transformations, the simplified effective energy function can be expressed as
\begin{align}
 E
 &= \sum_{i=1}^N (A_i - A_0)^2 + k_P (P_i - P_0)^2 + \Lambda L_i^{(n)}.
 \label{eq:energy}
\end{align}

%
%

The mechanical interaction force experienced by cell $i$ is defined as the gradient of Eq.~\ref{eq:energy}, i.e. $\mathbf{F_i}=-\mathbf{\nabla_i} E$. In addition to $\mathbf{F_i}$, we implement polarized self-propulsion force on each cell. Following the Self-Propelled Voronoi model~\cite{Bi_PRX_2016,huang_shear_2022}, we assign to cell $i$ a polarity vector $\mathbf{\hat{n}_i}$. The cell $i$ feels a self-propulsion force of constant magnitude $v_0/\mu$ directed along $\mathbf{\hat{n}_i}$, where the mobility $\mu$ is the inverse of a frictional drag. Taken together, these forces control the over-damped equation of motion of each cell center
\begin{align}
\frac{\mathrm{d} \mathbf{r_i}}{\mathrm{d} t} = \mu\mathbf{F_i} + v_0 \mathbf{\hat{n}_i}.
\end{align}
In actual biological systems, the polarity can depend on a variety of cues from a cell's past history and from interactions with neighbors~\cite{camley}. Here we adopt a simplified approach and let the polarity orientation $\mathbf{\hat{n}_i} = (\cos\theta_i, \sin\theta_i)$ obey rotational diffusion, given by 
\begin{align}
\partial \theta_i = \eta_i(t), \quad
\left\langle \eta_i(t) \eta_j(t')\right\rangle = 2D_r \delta(t-t')\delta_{ij},
\label{eq:abp}
\end{align}
where $\eta_i(t)$ is a white noise process with zero mean and variance $2D_r$. The magnitude of angular diffusivity $D_r$ determines the memory of stochastic noise of the self-propulsion direction, leading to a polarity persistence time scale $\tau=1/D_r$. When $D_r$ is small, the polarity direction $\mathbf{\hat{n}_i}$ changes slowly and leads to more persistent self-propelling force. For large $D_r\gg 1$, the corresponding persistence time scale $\tau\to 0$ is much shorter than other dynamical time scales in the system, and Eq. \eqref{eq:abp} approaches simple memory-free Brownian motion. In the SPV model, it has been observed that increasing $D_r$ can significantly decrease tissue fluidity at a fixed cell motility, highlighted by the finding that a solid-like tissue at large $D_r$ can be fluidized simply by reducing its $D_r$ value~\cite{Bi_PRX_2016,Yang_PNAS_2017}.

To study cell dynamics at medium density, we simulate a constant number ($N=400$) of cells under periodic boundary condition with a box size $L$ such that the packing fraction equals $\phi=0.5=N \pi/L^2$. The advantage of this choice compared to open boundary condition is that it conserves the density of self-propelled cells even with high diffusivity, and simulates scenarios where cells freely leave and enter the field of interest while the overall cell count is relatively stable. In this paper, the results shown correspond to $k_P=1$, $D_r=0.1$, and $\Lambda=0.2$ which reflects a relatively small difference between contacting and non-contacting edge tensions, unless otherwise specified. We choose $A_0=\pi$ to reflect the preferred area for an isolated cell. Changes of these parameters do not qualitatively change the phase diagram but merely shift the locations of phase boundaries. We initialize systems as one large connected cluster and relax it to steady state at zero temperature with $P_0$ ranging from $4$ to $8.8$. We then turn on $v_0$ with values ranging from $0$ to $2.7$ and conduct numerical simulations using molecular dynamics. We typically perform $10^6$ integration steps with a step size $\mathrm{d}t=0.01$ using Euler's method.



\section{Motility and cell-cell interactions induce a clustering transition}

\begin{figure*}
 \centering
 \includegraphics[width=0.8\textwidth]{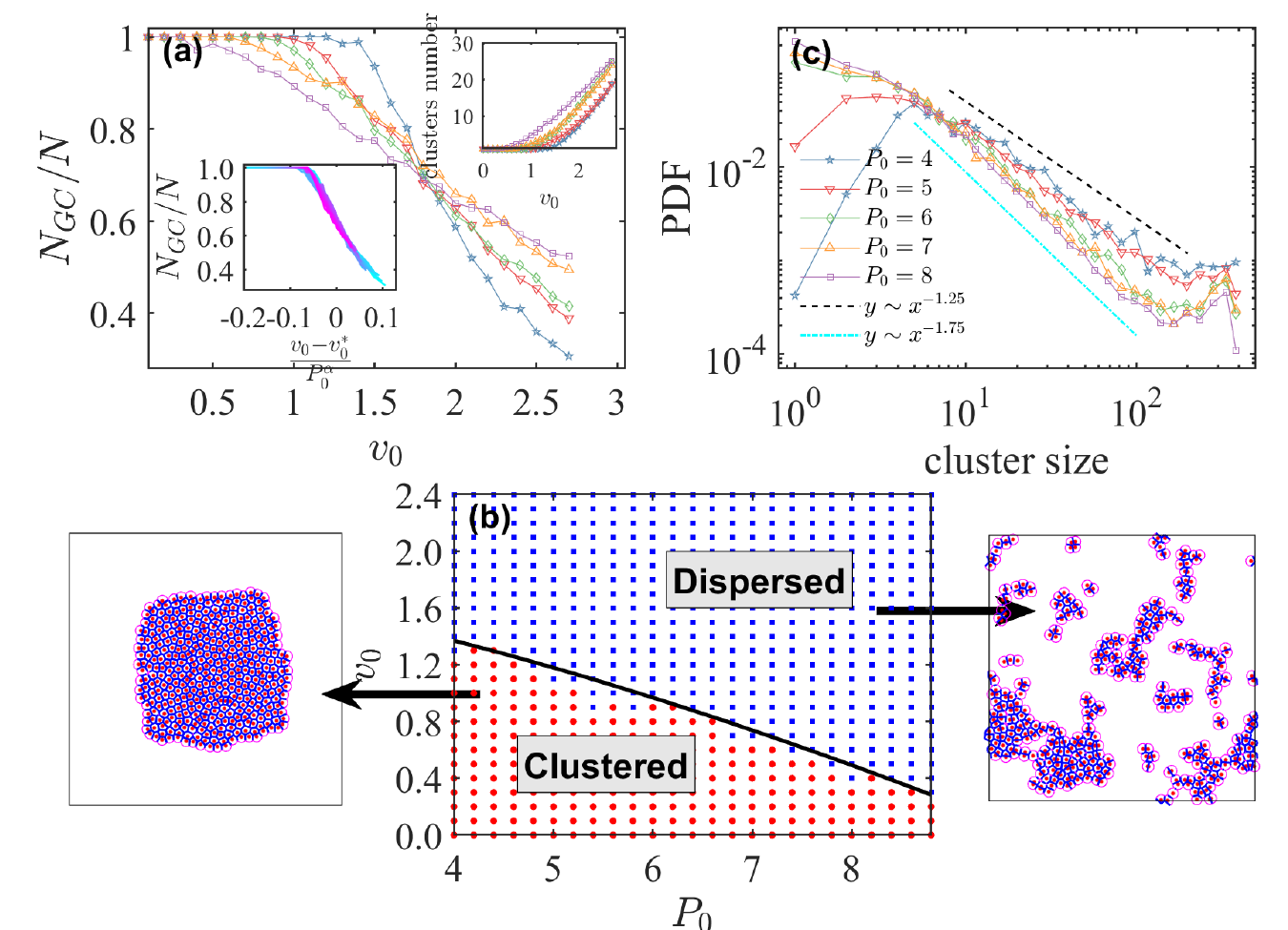}
	\caption{
 {\bf Formation of multicellular clusters in the AFV model.}
 (a) The normalized giant cluster size $N_{GC}/N$ as a function of $v_0$ at different $P_0$ and at $\Lambda=0.2$. An initially connected tissue undergoes a dispersion transition at a critical activity $v_s(P_0)$. Here the colors correspond to the legend in panel (b). (a-left-inset) $N_{GC}/N$ vs $(v_0-v_0^*)/P_0^\alpha$ for different $P_0$ values, which are indicated by the curve colors. (a-right-inset) The number of contiguous clusters as a function of $v_0$ for different $P_0$. (b) Tissue clustering phase diagram in the $v_0$-$P_0$ plane for $\Lambda=0.2$. The red data points correspond to the clustered state, and the blue points correspond to dispersed tissues. The black line corresponds to $v_s(P_0)$. The two sample tissue snapshots are from $P_0=4.2, v_0=0.1$ (clustered state) and $P_0=4.2,v_0=2.1$ (dispersed state). The red dots in the snapshots are cell centers, the blue lines are contacting edges between adjacent cells, and the pink arcs are the non-contacting edges. The black rectangles annotate the periodic boundaries of our computational domain. (c) The probability distribution function (PDF) for the cluster size for $N=400$ systems is shown at $v_0=v_0^*$. The dashed line and dot dashed line correspond to power-laws with exponents respectively of  $-1.25$ and $-1.75$, as visual aids. }
	\label{fig:cluster}
\end{figure*}

Migrating cells transition between dispersed individual and clustered multi-cellular collectives during embryonic development, tumor progression, and wound healing~\cite{Leggett_Neronha_Bhaskar_Sim_Perdikari_Wong_2019}. Although the contribution of these transitions to motility and coordinated behaviors are well-characterized for a confluent scenario, the behaviors for a non-confluent low-density scenario is poorly understood. 
As vertex-based models \cite{Alt_VM_review_2017} focus on scenarios where tightly packed cells cover the entire surface and form confluent epithelial monolayers, the ability of these models to simulate how cell connectivity affect global dynamics and tissue properties are limited. On the other hand, a previous study using the FV model \cite{Teomy_Kessler_Levine_2018} has not analyzed the structural organization of cell clusters over time nor included active forcing. These are clearly important aspects of tissue behavior.

Here, we characterize cell clustering by first identifying the ``giant cluster''~\cite{meder2006phase}, which is the largest connected cluster within the cellular system, and then analyzing the normalized giant cluster size, defined as $$\frac{N_{GC}}{N}\equiv\frac{\text{number of cells in the giant cluster}}{\text{total cell number }N} \in(0, 1],$$ which serves as an order parameter for cluster formation. For any instantaneous configuration, we consider two cells to belong to the same contiguous cluster if they share a contacting edge (presented as blue straight edges in Fig.~\ref{fig:sample}), or equivalently, when the distance between their centers is less than $2l$. 
%
%
In Fig. \ref{fig:cluster}(a), we plot the $N_{GC}/N$ as a function of the motility parameter $v_0$, for tissues at various values of $P_0$ and fixed $\Lambda=0.2$. A motility-driven dispersal transition is observed: at low $v_0$, cells form a single giant cluster corresponding to $N_{GC} = N$. Larger $v_0$ values cause a break-up into smaller clusters, indicated by a decreasing $N_{GC}/N$ value. We observe that the dispersal transition occurs for all $P_0$ values tested, where the sharpness of the transition depends on $P_0$. Intriguingly, all curves intersect at a common point $v_0^*\approx 1.85$, which serves as a  crossover independent of $P_0$. For $v_0<v_0^*$, $N_{GC}/N$ decreases as $P_0$ increases; when $v_0>v_0^*$, the behavior is flipped and $N_{GC}/N$ increases with $P_0$. Based on these observations, we hypothesize that the behavior of $N_{GC}/N$ below and above $v_0^*$ can be described by a  scaling relation
\begin{equation}
 N_{GC}/N = G\left(\frac{v_0-v_0^*}{P_0^\alpha}\right),
\end{equation}
where $G$ denotes a common scaling function. We replot all data in terms of $N_{GC}/N$ vs $\frac{v_0-v_0^*}{P_0^\alpha}$ based on the above ansatz in Fig. \ref{fig:cluster}(a) bottom-left inset, and obtain a collapse to a master curve with $v_0^*=1.85\pm0.06$ and $\alpha=1.50\pm0.12$. The existence of this collapse suggests that the transition will always occur at a motility threshold
\begin{equation}
 v_s(P_0)=v_0^*-c P_0^\alpha, 
 \label{v_s_boundary}
\end{equation}
where $c=0.06\pm0.005$ is a positive constant obtained through fitting to the location where $N_{GC}/N$ drops below 99\%. In addition, we plot the number of clusters as function of motility $v_0$ in Fig. \ref{fig:cluster}(a) top-right inset. The curves confirm the above observations that at low motility regime, cells remain in a large cluster, and the cluster number remains close to $1$. At larger $v_0$, large clusters start to break down into smaller pieces, indicated by an increasing cluster number. Then Eq.~\eqref{v_s_boundary} is shown in Fig.~\ref{fig:cluster}(b) which serves as a phase boundary between clustered and dispersed states. The clustered region is characterized by a high $N_{GC}/N$ value and low cluster number. When the dispersal transition point is exceeded, $N_{GC}/N$ starts to drop and the number of contiguous clusters increases, indicating entry into the dispersed region. In practice, we label tissues with $N_{GC}/N<0.99$ as being in the dispersed state indicated by blue dots, and those with $N_{GC}/N\le 0.99$ as being in a clustering state indicated by red squares. We also have included representative snapshots of the two states. 
%


The crossover point reveals an interesting regime in the phase diagram ($v_0 = v_0^* \approx 1.85$) where the system always possesses a giant cluster of a fixed size ($N_{GC}/N \approx 0.7$) regardless of the value of $P_0$. Given that this point occurs at a fairly large value of the motility, it leads to the question of how a large cluster size is maintained. We therefore analyzed the cluster size distribution (CSD) $p(n)$ for different $P_0$ values at the crossover point $v_0=v_0^*$. Fig. \ref{fig:cluster}(c) shows that the functional form of the CSD is dependent on $P_0$. 
At $P_0=4$, the CSD initially increases as a power-law of $n$, and then decreases at medium cluster sizes with a power law decay of $n^{-1.25}$ before finally reaching $N_{GC}$.
When $P_0=8$, the power-law decay has the form $n^{-1.75}$, suggesting a faster decay as cluster size increases. Similar exponents have been found in other active matter systems with clustering or motility induced phase separation~\cite{Huepe_Aldana_2004, Yang_Marceau_Gompper_2010, Linden_Alexander_Aarts_Dauchot_2019, Levis_Berthier_2014, Redner_Wagner_Baskaran_Hagan_2016, Zaccarelli_2007}. The broadness of the CSD suggests that these systems do not have a typical cluster size, indicating the need for a multi-scale analysis.

\begin{figure}
\centering
\includegraphics[width=0.5\textwidth]{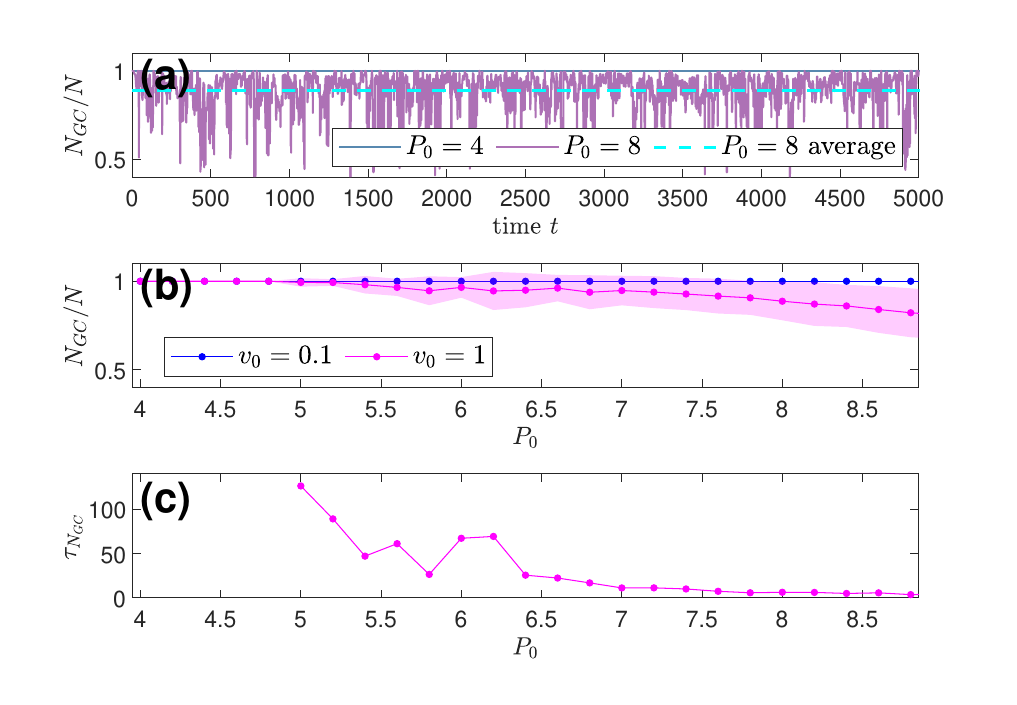}
	\caption{The fluctuation in $N_{GC}/N$ for dispersed states. (a) Two sample $N_{GC}/N$ traces at $v_0=1$. The solid lines correspond to realizations of $P_0=4$ and $P_0=8$, respectively. The cyan dashed line indicate the average value of $P_0=8$ as visual aids. (b) The $N_{GC}/N$ average values of different $P_0$ at $v_0=0.1$ and $v_0=1$, respectively, with the shaded band representing the standard deviation. (c) The autocorrelation time of the giant cluster size, $\tau_{N_{GC}}$, at different $P_0$ and $v_0=1$. For $P_0<5$, the $\tau_{N_{GC}}$ values go to infinity and are not displayed on the plot.}
	\label{fig:gcs_trace}
\end{figure}

Having quantified the steady-state mean size of the cluster, we now shift our focus to the temporal fluctuations in cluster sizes. As illustrated in the simulation videos provided in the Supplementary Materials, cell clusters can exhibit significant dynamism. Over time, cells may aggregate and disintegrate from a cluster, reflecting a highly dynamic process. The temporal fluctuation of $N_{GC}/N$ for two representative cases is shown in Fig.~\ref{fig:gcs_trace}(a).  For a clustered state ($v_0 = 1$ and $P_0 = 4$), where the giant cluster is stable, the $N_{GC}/N$ trace is a flat line equals to $1$. On the other hand, as shown in Fig. \ref{fig:gcs_trace}(a,b), in a dispersed state, the giant cluster size experiences continuous fluctuations as clusters dynamically form and breakup. As cells interact and coalesce, larger clusters emerge, resulting in an increase in the giant cluster size. However, the  instability of cell contacts in such a state causes these clusters to be inherently transient. Over time, the clusters disintegrate, leading to a reduction in the giant cluster size. We also investigate the temporal evolution of cluster sizes with using the autocorrelation time $\tau_{N_{GC}}$ of the giant cluster size. Fig. \ref{fig:gcs_trace}(c) shows  $\tau_{N_{GC}}$ for $v_0=1$, when $P_0<5$ the giant cluster size $N_{GC}$ is always equal to $N$ and the autocorrelation time is infinity. When $P_0>5$ the system enters dispersed state, and as $P_0$ increases the autocorrelation time decreases, indicative of a tendency towards unstable cluster and faster cluster size fluctuations.

\section{Epithelial-Mesenchymal Transition Driven by Cell-Shape Changes and Motility}

\begin{figure}
\centering
\includegraphics[width=0.48\textwidth]{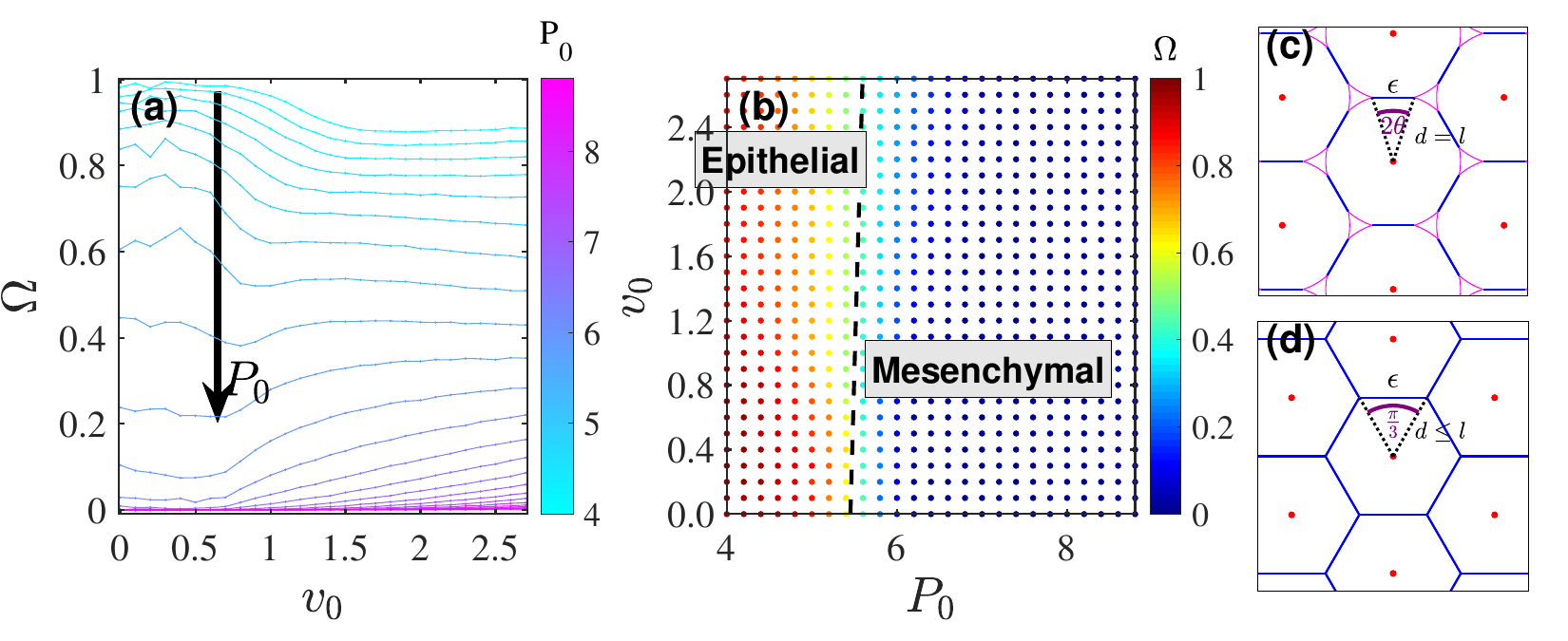}
	\caption{Tissue confluency properties. (a) $\Omega$ [defined in Eq. \eqref{eq:omega}] as a function of $v_0$ for various $P_0$ values ranging from $4$ to $8.8$ with an increment of $0.2$. The curve color corresponds to the tissue $P_0$ value. The black arrow indicates the curve ordering with increasing $P_0$. (b) Tissue confluency phase diagram for $\Lambda=0.2$ in the $v_0$-$P_0$ plane. The marker color indicates the tissue $\Omega$ value. The black dashed line corresponds to $\Omega=0.5$ and is used as a boundary between epithelial/confluent state and mesenchymal/nonconfluent state. (c-d) Illustrative snapshots of a nonconfluent/confluent hexagonal tissue, respectively, used for our mean-field analysis. }
	\label{fig:confluent}
\end{figure}



In a confluent epithelial layer, cells pack without intercellular gaps, and triple junctions exist between any trio of three neighboring cells. On the other hand, a classical signature of EMT is the loosening of tight epithelial tissue organization. This is accomplished by transcriptional suppression of E-cadherin, a prototypical adhesion molecule responsible for maintaining cell-cell adherens junctions, as well as the downregulation of other adhesion molecules such as claudins~\cite{thiery2009epithelial, lamouille2014molecular}. Indeed, tissues undergoing EMT exhibit decreased tight junctions \cite{emt-junctions1,emt-junctions2} and increased intercellular gaps \cite{mitchel_ncomm_2020}; in the developmental biology context, these changes can be imaged in vivo~\cite{amack2021cellular}. In other words, when cells lose their epithelial character as they become mesenchymal,  triple junction vertices will be lost and gradually replaced by the presence of intercellular gaps~\cite{leggett2016morphological, mitchel_ncomm_2020}. This morphology change reduces tensions transmitted at cell-cell junctions and eventually helps enable cell movement~\cite{nunan2015ephrin, nieto2016emt}. 

Based on these observations, we searched for an order parameter that could reflect the ratio between triple junction vertices and inter-cellular gaps in order to quantify the degree of ``epithelial-ness'' of a cell layer. Euler's polyhedron formula asserts that for a completely confluent cluster of $N^{(j)}$ cells in open space, the number of triple junction vertices $V_3^{(j)}$ equals to $E_c^{(j)}-N^{(j)}+1$, where $E_c^{(j)}$ is the number of contacting edges. When inter-cellular gaps develop, $V_3^{(j)}$ will decrease and deviate from $E_c^{(j)}-N^{(j)}+1$. Thus, a natural confluency order parameter for cell cluster $j$ can be defined as
\begin{equation}
 \omega^{(j)} = \frac{V_3^{(j)}}{max(E_c^{(j)}-N^{(j)}+1,\, 1)} \in [0, 1],
\end{equation}
where the denominator has a lower bound of $1$ to avoid dividing by $0$ when the cluster is made of a linear string of cells. The order parameter $\omega^{(j)}=0$ for nonconfluent clusters without any triple junction vertex, and $\omega^{(j)}=1$ for confluent clusters. Then, for a system containing multiple contiguous clusters, the global confluency order parameter can be defined as weighted average of $\omega^{(j)}$
\begin{equation}
 \Omega = \frac{\sum_j N^{(j)} \omega^{(j)}}{\sum_j N^{(j)}}.
 \label{eq:omega}
\end{equation}

The behavior of $\Omega$ as a function of $P_0$ and $v_0$ is shown in Fig. \ref{fig:confluent}(a). When $P_0$ is low, cells are close-packed with each other, resulting in confluent epithelial tissues with $\Omega \sim 1$. As $P_0$ increases, tissue confluency is gradually lost, indicated by a decreasing $\Omega$. The cell motility $v_0$ has a weak yet opposite effect on $\Omega$ for systems with different $P_0$; cell motility enhances inter-cellular gaps at low $P_0$ while it tends to eliminate inter-cellular gaps at high $P_0$. In Fig. \ref{fig:confluent}(b), we use $\Omega=0.5$ as the threshold to distinguish an epithelial state, where triple junctions are formed at a dominant fraction of the cell-cell interface inside clusters, and a mesenchymal state, where intercellular gaps are preferred. The E/M phase boundary is nearly vertical, indicating a strong dependence on $P_0$.

To better understand the nature of the epithelial to mesenchymal transition in this model, we utilize a simple mean-field calculation at zero motility. 
Consider a simple case of a hexagonal cell packing. As illustrated in Fig. \ref{fig:confluent}(c-d), each cell has exactly 6 neighbors whose centers are located on the vertices of a regular honeycomb lattice whose center coincides with the central cell center. Each contacting edge has length $\epsilon$, and the corresponding central angle equals $2\theta$ where $\epsilon=2\sin\theta$. When $\theta<\pi/6$, inter-cellular gaps exist instead of triple junction vertices. As the non-contacting perimeter and area of each cell are given by
\begin{eqnarray}
 L^{(n)}&=&2\pi - 12\theta, \\
 A&=& (\pi - 6\theta) + 3\sin\theta\cos\theta,
\end{eqnarray}
we can determine the mechanically stable configuration by finding the minimum of Eq. \eqref{eq:energy} with respect to the angle $\theta$. A simple calculation yields
\begin{align}
 0 = \frac1N \frac{\partial E}{\partial \theta} 
 =~& 24 k_P(1-\cos\theta)(12\theta-12\sin\theta+P_0-2\pi) \notag \\
 &+ 24\sin^2\theta(6\theta-3\sin 2\theta + A_0 -2\pi)-12\Lambda.
 \label{confluency_condition}
\end{align}
The confluency transition is predicted to occur when the solution to Eq.~\eqref{confluency_condition} is given by $\theta=\pi/6$. For the parameter set used in this paper ($\Lambda = 0.2,~ A_0 = \pi,~ k_P = 1$), this condition gives the critical point of $P_0=5.73$, below which the system is confluent. This estimate provides a reasonable approximation for the phase boundary between epithelial and mesenchymal states in Fig.~\ref{fig:confluent}(b), in the limit of $v_0 \to 0$. Nevertheless, this mean-field analysis is a simplification as it captures the EMT as a clear-cut switch rather than a spectrum.

\section{Glassy dynamics in confluent and non-confluent regimes}

So far we have characterized the behavior of the model based on static, structural properties of the multicellular organization. Next, we analyze the dynamical behavior of the model. Previous studies have suggested that cells have the ability to transition from solid-like to fluid-like state via separate pathways. For example, during UJT, a tissue can fluidize while remaining confluent~\cite{Park_NMAT_2015,atia2018geometric,malinverno2017endocytic}. While during EMT \cite{nieto2016emt, mitchel_ncomm_2020}, abolition of the epithelial character is a necessary precondition for promoting cell migration. Here, a natural question arises; while dispersed states must be fluid-like, are all clustered phases solid-like? Further, do the structural transitions observed in our model coincide with dynamical transitions?


\begin{figure}
\centering
\includegraphics[width=0.5\textwidth]{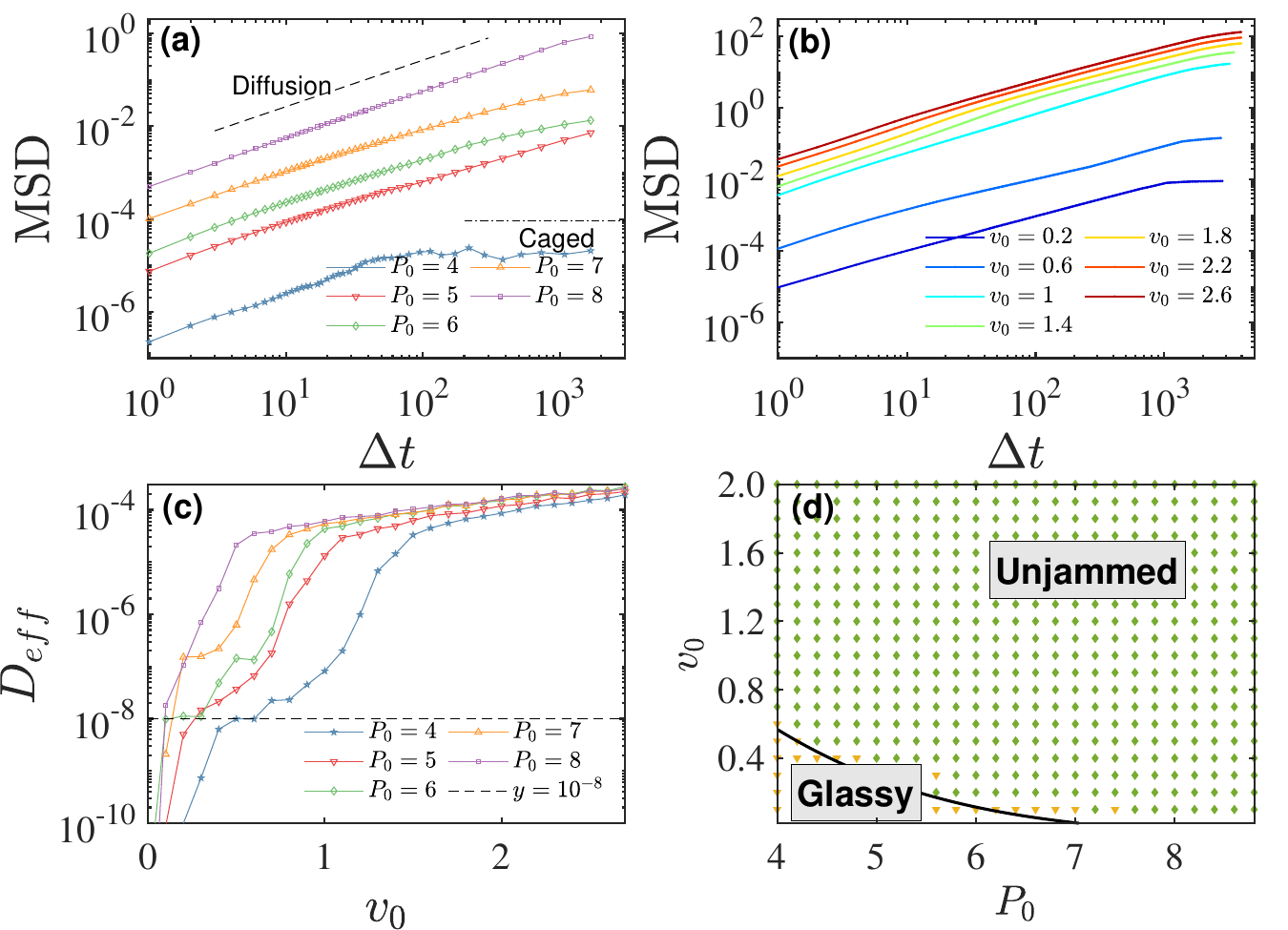}
	\caption{Tissue glassy properties. (a) MSD traces for different $P_0$ ranging from $4$ to $8$, with $\Lambda=0.2$ and $v_0=0.4$.
	(b) MSD traces for different $v_0$ with a fixed $P_0=5.6$.
	(c) $D_{eff}$ as a function of $v_0$ for different $P_0$. The black dashed line corresponds to $D_{eff}=10^{-8}$. (d) Tissue diffusivity phase diagram as a function of $P_0$ and $v_0$. Orange data points correspond to glassy tissues with vanishing $D_{eff}$; green points correspond to flowing tissues (finite $D_{eff}$).
}
	\label{fig:diffusion}
\end{figure}

	
	
\begin{figure}
\centering
\includegraphics[width=0.4\textwidth]{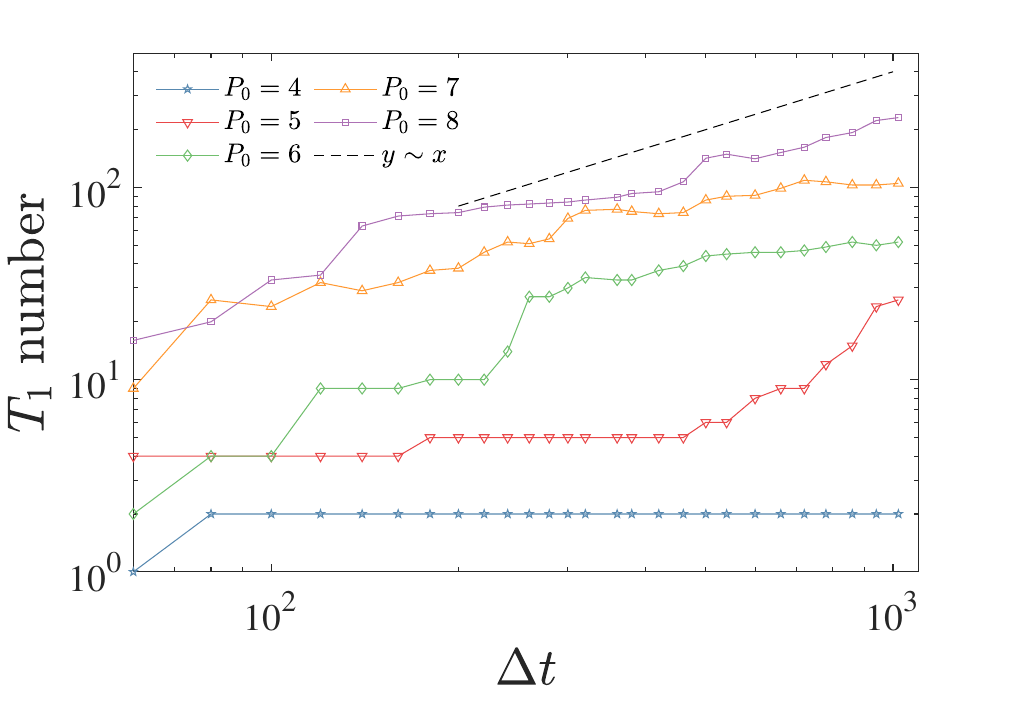}
	\caption{The cumulative T1 number traces for different $P_0$ ranging from $4$ to $8$ with $\Lambda=0.2$ and $v_0=0.4$.}
	\label{fig:t1}
\end{figure}

To answer this question, we characterize the dynamics of cell motion in our system by measuring the mean-squared displacement (MSD). In order to exclude the contribution of collective rigid-body translations and rigid rotations of  clusters, 
we compute the MSD based on the relative displacement of cells with respect to their nearest-neighbors. In Fig. \ref{fig:diffusion}(ab), we plot the MSD as a function of time lag $\Delta t$ for systems with different $P_0$ and $v_0$ values. For small $v_0$ values, cells are caged at long time scales, as indicated by the plateau of MSD. We also plot the total number of $T_1$ transitions in Fig. \ref{fig:t1}, suggesting arrested motility due to caging effects and broken ergodicity, both of which are characteristic signatures of glassy dynamics. As $v_0$ increases, cells begin to uncage and the MSD increases asymptotically as $MSD\propto \Delta t^\beta$, where $\beta\sim 0$ for $P_0\le 4$ and $\beta\sim1$ for $P_0\ge 7$.
	
Following previous studies on tissue glassy dynamics \cite{Bi_PRX_2016}, we use the self-diffusivity $D_s=\lim_{\Delta t\to\infty} MSD(\Delta t)/(4\Delta t)$ as an order parameter to distinguish glassy and fluid states. We ran simulations for $10^{4}$ time units and used $\Delta t=5000$ to calculate $D_s$, which is much longer than the typical caging time scale in fluid state. The calculated $D_s$ is presented in units of $D_0=v_0^2/(2D_r)$, the free diffusion constant of a self-propelling cell, to accommodate the effect of varying motility. Then, the effective diffusivity $D_{eff}\equiv D_s/D_0$ is used as an order parameter to distinguish glassy (jammed) and fluid (unjammed) states. The behavior of $D_{eff}$ at different $P_0$ and $v_0$ is shown in Fig. \ref{fig:diffusion}(c). For a given low $v_0$ value, the order parameter $D_{eff}$ does not necessarily follow the ordering of $P_0$ values; At large $v_0$ regime, high $P_0$ systems always correspond to high $D_{eff}$ values. In Fig. \ref{fig:diffusion}(d), we plot phase diagram of cell dynamics in the $v_0$-$P_0$ plane according to $D_{eff}$: The glassy states correspond to a finite $D_{eff}$ below a noise floor of $10^{-8}$, and the unjammed states correspond to $D_{eff}$ that exceeds this threshold.

The position of the dynamical phase boundary suggests that the energy barrier for cell rearrangements is lower than that for cluster breakup. This difference gives raise to the existence of stable fluid-like clusters, within which cells exchange neighbors frequently yet stay as members of the same connected cluster. This possibility is in good agreement with experimental observations of bulk epithelial colonies, for example Madin-Darby canine kidney (MDCK) cells form a confluent epithelial sheet through a highly motile expanding process lasting for one week~\cite{Puliafito_Hufnagel_Neveu_Streichan_Sigal_Fygenson_Shraiman_2012, Heinrich_Alert_LaChance_Zajdel_Kosmrlj_Cohen_2020}.

Next we also use the number of cell rearrangements as an alternative indicator of tissue fluidity. To this end, we generalize the concept of a $T_1$ transition for use in the AFV model: In confluent vertex- or Voronoi-based models, a system of $N$ cells possesses $3N$ edges, and each cell rearrangement occurs as a $T_1$ event, which involves the elimination of an existent edge and formation of a new edge. As the total edge number is no longer conserved in the AFV model, edge elimination and new edge formation can happen independently. To generalize the method to count cell rearrangements, we consider the independent elimination or formation of an edge each as half of a $T_1$ event; this choice will give the usual $T_1$ transition number for confluent states~\cite{das_T1_prx}. In addition, the elimination and formation of a new edge between a given pair of cells will be taken to cancel out each other, and contribute $0$ to the $T1$ counting rather than $1$. This choice is necessary because cell contacts are ``shallow'' in some jammed mesenchymal systems, where a pair of cells could keep forming and breaking a short contacting edge due to local fluctuations. In Fig. \ref{fig:t1}, we show some sample traces of the total $T_1$ rearrangement counted in such a manner. The $T_1$ traces confirm the existence of jammed/unjammed phases, as discussed in the main text. For systems in a jammed state, e.g. $P_0=4$ at $v_0=0.4$, the $T_1$ number trace suggests the cells are caged by their neighbors. On the other hand, for systems in an unjammed state, the $T_1$ rearrangement number increases over time, indicating a  diffusive behavior.

\section{Discussion and Conclusions}

\begin{figure*}
 \centering
 \includegraphics[height=8cm]{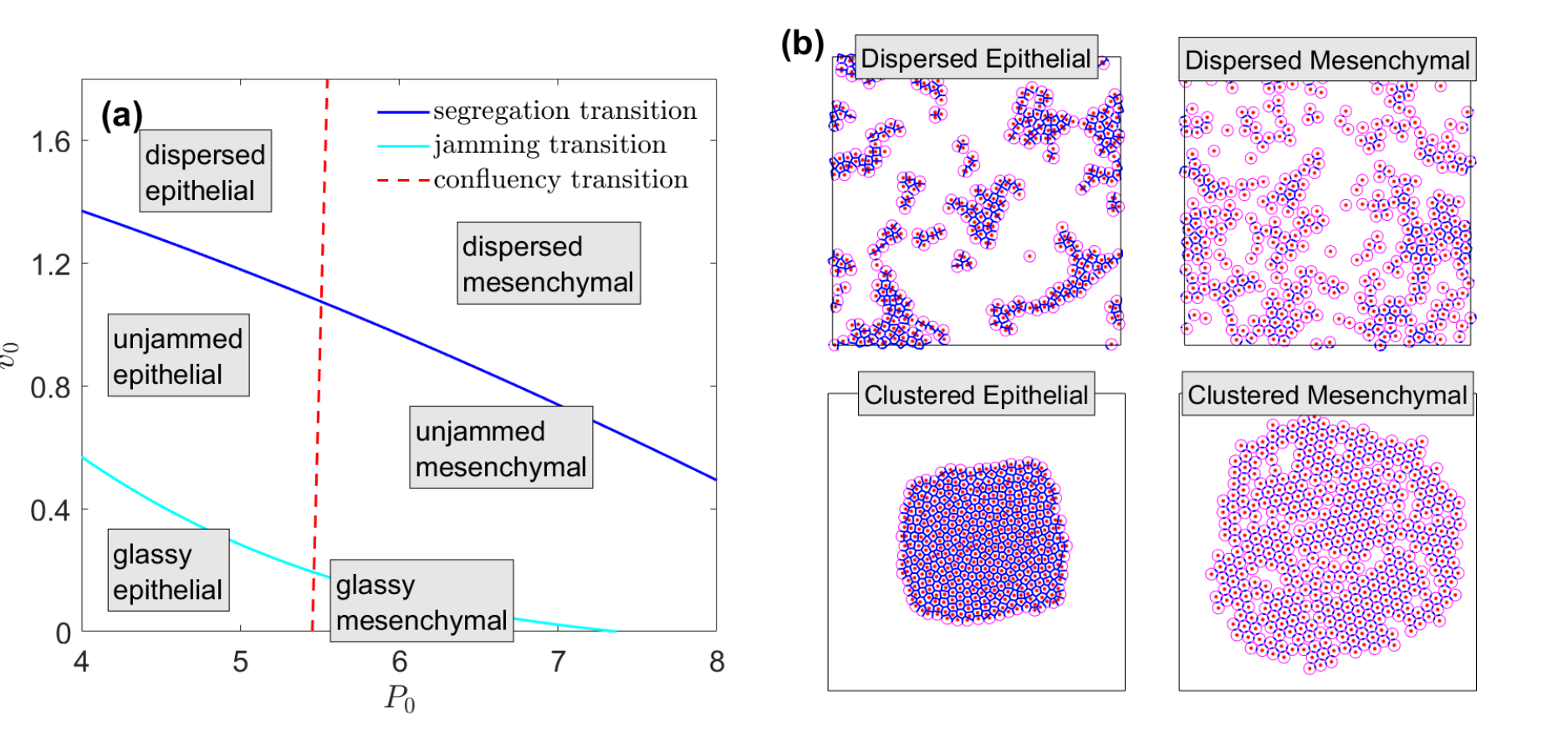}
	\caption{Summary of phase diagram. (a) Sketch of the phase diagram with all three phase boundaries, on the $v_0$-$P_0$ plane for $\Lambda=0.2$. This structure is qualitatively robust for all values of $\Lambda$. (b)  Sample snapshots of four dictinct phases. The $(P_0, v_0)$ values for each state are: Clustered Epithelial phase $(4, 0.1)$; Clustered Mesenchymal phase $(8.8, 0.1)$; Dispersed Epithelial phase $(4.4, 2.6)$; Dispersed Mesenchymal phase $(8.8, 2.6)$. Note that clustered state could be glassy or unjammed, but the dispersed state must be unjammed.}
	\label{fig:summary}
\end{figure*}

In this work, we have introduced a new approach to tissue dynamics, referred to as the Active Finite Voronoi (AFV) model. This model retains the simplicity of the active Voronoi/vertex-based approach, which has made it a very popular strategy for studying confluent epithelial systems. However, the addition of a maximal size for cells allows the system to dynamically determine the degree of confluency as a function of the system parameters. The observation of the E/M transition using our model extends the behavior of the vertex-based models studied previously. The traditional vertex model, constrained to be confluent, exhibits a transition to a fluid-like phase at high $P_0$, and there is only a confluent and fluid-like phase for high $P_0$ systems. In contrast, our new model allows for the observation of a new non-confluent phase wherein cells are given the possibility of developing intercellular gaps. Furthermore,  the confluency transition is not necessarily coincident with the tissue dispersal and glassy transitions. As we have seen, this enables the system to exhibit an epithelial-mesenchymal transition, as has been observed in developmental biology, wound healing, and cancer metastasis.

In general, the AFV exhibits three phase transitions and thereby defines six different possible phases, as shown in Fig. \ref{fig:summary}(a). One critical parameter is the preferred perimeter, $P_0$. On the low $P_0$ side, triple junction vertices are energetically preferred, indicating an epithelial state; On the high $P_0$ side, cell-cell interfaces are dominated by inter-cellular gaps, suggesting that tissues are in a mesenchymal state. Parallel to these two states, when $v_0$ is low, tissues stay in glassy (clustered and dynamically arrested) state; by increasing $v_0$, cell rearrangements become more frequent and tissues can enter unjammed (clustered and flowing) state; finally, when $v_0$ is high, cells are no longer able to stay connected, and tissues transition to  a dispersed state which much be flowing. We show some sample instantaneous tissues snapshots in Fig. \ref{fig:summary}(b). Sample videos of these six phases are included in Supplementary Materials.

In the AFV, isolated cells, whether fully disconnected or shallowly linked with others, by design, retain a round shape. This aspect of the model is a simplification, given that in experimental observations, cells devoid of epithelial connectivity often exhibit a more spindle-like form. For future exploration, it would be beneficial to incorporate a version of the AFV model that provides cells with additional degrees of freedom, such as elongation. This enhancement would permit cells to assume non-circular shapes when isolated.

In this work, we did not consider cell proliferation and apoptosis in our model. Should cell density increase overall due to cell division or apoptosis, we anticipate that the tissue would experience density-driven jamming~\cite{angelini_pnas}. However, it's also likely that a real tissue would manifest contact inhibition of locomotion~\cite{Mayor_CIL} as cell density rises, corresponding to a reduction in $v_0$ with increased density. Intriguingly, even in situations where tissue homeostasis is maintained through an equilibrium between cell division and apoptosis, the tissue would inevitably fluidize due to the constant injection of energy from cell division~\cite{tang2023cell}.
Incorporating changes in number density would introduce an additional dimension to the phase diagram of AFV. This could be an engaging avenue for future exploration. Concerning cell-matrix interactions, our model adopts a simplified approach, encapsulating only the viscous friction between the cell and the extracellular matrix or substrate. It could provide valuable insights to expand our analysis and incorporate a more comprehensive model for the viscoelastic cell-matrix interactions, possibly paralleling the approach employed by Ajeti et. al~\cite{ajeti2019wound}.

It is worth mentioning that all of these phase transitions are reversible. For example, by changing $v_0$ from $0.2$ to $2$, systems with $P_0=4$ undergo a dispersal transition characterized by breaking apart of the bulk. Once the $v_0$ value is reset to $0.2$, small contiguous clusters gradually merge into bigger clusters, once they collide during drifting. Given a long enough time, the system is always able to revert back to the clustered state, even though the final merged clusters have some morphological differences from the original ones; for example, there are more holes inside the clusters, and the contour shapes are more irregular. Thus, there can be some level of microscopic hysteresis, but none at the level of the macroscopic phase structure.
	
There are a variety of experimental systems that can be studied with this new model. Wong's group has demonstrated \cite{Leggett_Neronha_Bhaskar_Sim_Perdikari_Wong_2019} that under the right conditions cells can form disconnected fractal clusters, similar in principle to those illustrated in Fig. \ref{fig:sample}. In a simple animal {\it (Trichoplax adhaerens \cite{schierwater}}), overall motility can introduce enough stress to cause fracture of the epithelial tissue; amazingly, the fracture can transition from brittle to ductile behavior~\cite{prakash}. Finally, the issue of the detachment of cell clusters from primary tumors is very much at the heart of trying to understand the initial stages of the metastatic process~\cite{pathology,budding,fuhs2022rigid}. The AFV can be used to predict cluster size distributions and thereby provide a check on the accuracy of previous attempts \cite{mukherjee2021cluster} to accomplish this task.

\section*{Acknowledgements}
We thank David Kessler for useful discussions. This work was supported in part by NSF DMR-2046683 (J. H. and D. B.), PHY-1935762 (H. L.) and PHY-2019745 (J. H., D.B., and H. L.). We also acknowledge support provided by the Alfred P. Sloan Foundation (J. H. and D. B.) and the Human Frontier Science Program (J. H. and D. B.).

\bibliography{main.bib}
\bibliographystyle{rsc}

\cleardoublepage

\end{document}